# Intelligent Electromagnetic Sensing with Learnable Data Acquisition and Processing

Hao-Yang Li[1], Han-Ting Zhao[1], Meng-Lin Wei[1], Heng-Xin Ruan[1], Ya Shuang[1], Tie Jun Cui[2], Lianlin Li[1]

State Key Laboratory of Advanced Optical Communication Systems and Networks, Department of Electronics, Peking University, Beijing 100871, China

[2] State Key Laboratory of Millimeter Waves, Southeast University, Nanjing 210096, China

**Electromagnetic (EM) sensing is a wide-spread contactless examination technique in science, engineering and military. However, conventional sensing systems are mostly lack of intelligence, which not only require expensive hardware and complicated computational algorithms, but also pose important challenges for advanced in-situ sensing. To address this shortcoming, we propose the concept of intelligent sensing by designing a programmable metasurface for data-driven learnable data acquisition, and integrating it into a data-driven learnable data processing pipeline. This strategy allows to learn an optimal sensing chain in systematic sense of variational autoencoder, i.e., to jointly learn an optimal measurement strategy along with matching data post processing schemes. A three-port deep artificial neural network (ANN) is designed to characterize the measurement process, such that an optimal measurement strategy is adaptive to the subject of interest by controlling the programmable metasurface for manipulating the EM illuminations. We design and fabricate a proof-of-principle sensing system in microwave, and demonstrate experimentally its significance on the high-quality imaging and high-accuracy object recognition from a remarkably reduced number of measurements. We faithfully expect that the presented methodology will provide us with a fundamentally new perspective on the design of intelligent sensing architectures at various frequencies, and beyond.**

## Introduction

Electromagnetic (EM) sensing is a widely-used contactless examination technique owing to its unique properties of harmlessness, cost efficiency, portability and speed, which has been playing a critical role in various applications such as medical diagnostic[1-3], safety screening[4-6], earth resource exploration[7-9], and so on[10-11]. However, the conventional EM sensing systems are mostly lack of intelligence, which not only require considerable expenses on hardware and computational algorithms, but also pose important challenges for advanced in-situ sensing in complex surrounding environment. Here, we mean by intelligence that data acquisition and processing pipeline, working as biological systems, are learnable, scene adaptive and specific-task adaptive. In many cases like health monitoring or security screening, a measurement setting, for instance, a microwave radar, is conventionally not tailored to the specific task, instead, the hardware is designed toward generating a full-pixel high-resolution image as a crucial checkpoint for subsequent tasks. However, especially in the era with strong trends towards three-dimensional (3D) and high-frame-rate imaging, the flux of sensing data is so large that full-resolution imaging is very inefficient, and is also a big waste since usually only a few properties of the images are interested in practice, such as 'where is a particular object in the scene?' or/and 'what is it doing?' etc. This kind of situations requires a sensor to instantly recognize the feature of interest and make real-time decision, i.e. rendering the important features with high speed, fidelity, and compression ratio.

There are rapidly growing interests in applying machine-learning (ML) techniques to improve the EM sensing performance by developing data-driven learnable strategies[12-18]. For instance, the deep learning solutions have been developed for efficiently recovering an image of scene or accurately recognizing embedded features from a highly reduced number of measurements[12-15]. In these cases, just a few relevant data are physically recorded to register the scene without losing the information of interest, and thereby relieve remarkably the big pressure on storage, transfer and processing of full-pixel original scenes. Recently, we have demonstrated that deep learning techniques can provide considerable impacts on data processing pipeline, especially when explicit analytical treatments are of lack or intractable for modelling undergoing signals and hardware systems[16]. Despite all these advances, current sensing systems remain lack of intelligence since their data acquisition and processing schemes are designed separately, where many knowledges being crucial for reducing the number of measurements and improving the sensing performance have been ignored, notably on the level of hardware design, as noted with aforementioned microwave radar examples. Thereby, the conventional sensing systems are leaned on either expensive hardware, or time-consuming data acquisition, or expensive computational efforts.

Data acquisition and data processing are two critical building components of an EM sensing chain, which should have worked as in an integrated system to achieve the optimal sensing performance with as low expense on hardware and computation as possible. To achieve the EM sensing with "intelligence", we make two-aspect contributions. First, we propose a data-driven learnable data acquisition scheme by designing a programmable metasurface, which is controlled with a deep artificial neural network (ANN) (called the *measurement ANN, m-ANN in short*). Second, we jointly optimize the control coding patterns on the metasurface together with another ANN (called the *reconstruction ANN, r-ANN in short*) for retrieving relevant scene information in sense of variational autoencoder (VAE). In this framework, the data acquisition as a trainable physical network has been integrated with the data processing pipeline into a whole sensing chain. By training the entire sensing chain with a standard supervised learning procedure, we can simultaneously determine an optimal measurement scheme, i.e., a set of optimal coding patterns of the metasurface for generating an optimal illumination scheme, for capturing relevant scene information on hardware level, along with a matching data processing scheme for extracting relevant information from measurements. With such a strategy, the optimal sensing performance with a drastically minimized number of measurements can be achieved, which can help us to remarkably improve many critical metrics, for instance, energy consumption, hardware cost, efficiency, and so on. As we will detail below, this framework is general, flexible, and will provide considerable impacts on future sensing scenarios.

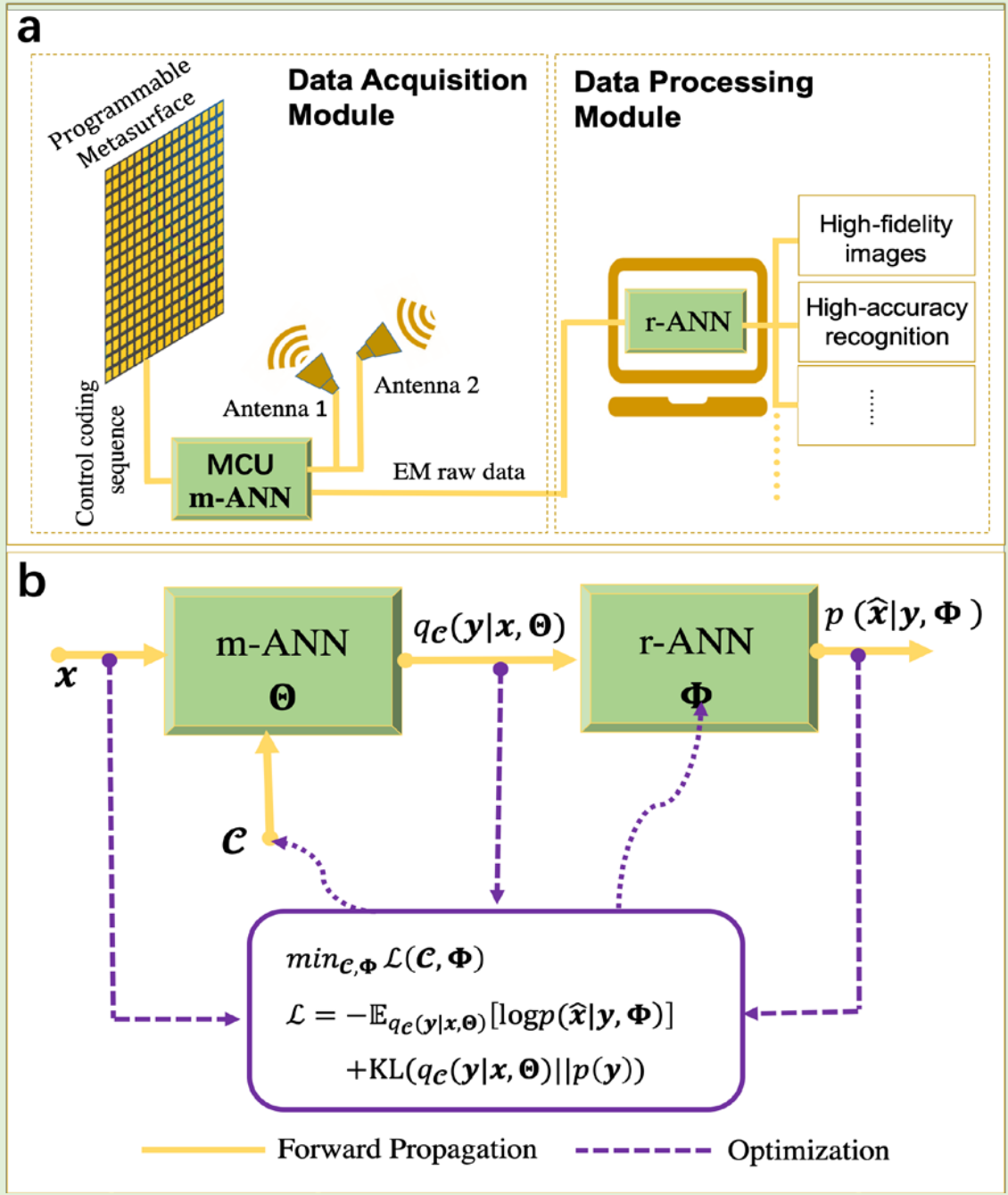

**Figure 1. Working principle of the intelligent sensing system.** (a) The intelligent sensing system consists of two data-driven learnable modules, i.e., the m-ANN-driven data acquisition module, and the r-ANN-driven data processing module. For the data acquisition module, it consists of a pair of horn antennas and an m-ANN-controlled programmable coding metasurface. The operational principle of the presented sensing system is described as following. Antenna 1 is used to emit periodically microwave illumination signals, which will be shaped by the programmable metasurface. After being scattered by the subject of interest, the wavefields shaped by the metasurface are received by Antenna 2. Finally, the received microwave raw data are instantly processed by the r-ANN, producing the desired imaging or recognition results. (**b**) the m-ANN-based data acquisition as a trainable physical network is fully integrated with the r-ANN-driven data processing pipeline into a whole sensing chain. Note that the m-ANN has three ports: one is used for receiving the scene of interest $x$, one is for inputting the control coding pattern $c$ of the metasurface for generating the optimal illumination, and the other is for outputting the EM raw data $y$. An optimal setting of the entire sensing chain can be learned by using a standard supervised learning procedure, for instance, the back-propagation algorithms well developed in TensorFlow, in context of VAE. In this way, we can simultaneously determine the optimal measurement scheme for capturing relevant scene information on hardware level, along with a matching data processing scheme for extracting relevant information from measurements.

In this article, we develop a proof-of-principle intelligent sensing system in the microwave frequency, and experimentally show the significance of the proposed data-driven learnable sensing strategy on image quality and accuracy of the object recognition compared to the conventional counterparts. It can be faithfully expected that the presented methodology could provide us a fundamentally new perspective on the design of sensing architectures at various frequencies.

## Materials and Results

**System configuration and operational principle.** We here elaborate on the configuration of the proposed intelligent sensing system and its operational principle. For illustrative purpose, the intelligent sensing system is designed in the microwave frequency; however, it can be readily extended to other frequencies. With reference to **Figure 1a**, the presented intelligent sensing system is mainly composed of two data-driven learnable building modules, i.e., one is for adaptive data acquisition, and the other is for instant data processing. Notice that the learnable data acquisition is centered around an m-ANN-driven programmable metasurface. More details about the programmable metasurface can be found in **Methods** and **Supplementary Note 1**. The programmable metasurface can be trained on the physical level with deep learning techniques to generate the scene-adaptive illuminations. Moreover, the resultant measurements are consistent with those required by data processing schemes. To achieve an optimal measurement scheme by manipulating the coding pattern of the metasurface, the setting of data acquisition is modelled with a deep ANN in this work. This ANN is data-driven learnable, which can be trained with a standard supervised learning procedure to achieve an optimal measurement scheme for probing a specific subject. For this reason, we would like to refer it to as the measurement ANN (m-ANN in short). Unlike conventional end-to-end ANNs with two ports, the measurement ANN has three ports, as shown in **Figure 1b**, which are connected with the coding pattern of the metasurface, the probed scene, and the output complex-valued microwave raw data, respectively. The microwave raw data will be input to a data-driven learnable data processing module, which is realized by another ANN, for retrieving relevant scene information. For this reason, such an ANN is called the reconstruction ANN (r-ANN in short). The reconstruction ANN can also be trained with a standard supervised learning procedure. As such, we can simultaneously determine the optimal measurement strategy to encode relevant scene information, along with a matching data processing algorithm to decode relevant information from measurements. In a nutshell, the data acquisition process and data post processing pipeline are integrated as a data-driven learnable system, and such a sensing chain can be learned to be fully adaptive to both the scene of interest and the sensing hardware. For these reasons, we would like to claim the term "intelligence" for our sensing system.

Here, we provide an insight into the principle behind the presented intelligent sensing system in context of VAE[19-21], a widespread probabilistic ML approach. The entire intelligent sensing can be viewed as a user-controlled end-to-end process that, under the control of the metasurface configured with coding patterns $C$, a given scene denoted by $x$ will generate the measurements $y$ by sampling from a $\mathcal{C}$-controllable conditional distribution $y \sim q_\mathcal{C}(y|x, \Theta)$. This distribution is responsible for modelling the measurement process of proposed intelligent sensing system. The primary goal of data processing is to find an estimator for retrieving the relevant scene information $x$ from the measurements $y$. Clearly, the estimator serves as an inverse action of measurement process $y \sim q_\mathcal{C}(y|x, \Theta)$, which is modeled as a conditional distribution $x \sim p(x|y, \Phi)$, where $\Phi$ encapsulates all learnable weights of the r-ANN. As emphasized above, the optimal sensing performance can be achieved by simultaneously learning optimal measurement strategy and the optimal reconstruction scheme. The network weights $\Theta$ of the m-ANN can be readily learned from the $(x, \mathcal{C}, y)$-triple labeled training dataset by exploring a standard supervised learning procedure, as detailed in **Supplementary Note 4.** In context of VAE, the optimal settings of the control coding pattern $\mathcal{C}$ and the network weights $\Phi$ of the r-ANN can be learned in context of VAE. Specifically, they can be found by minimizing the following objective function, i.e.,

$$\mathcal{L}(\mathcal{C}, \Phi) = -\mathbb{E}_{q_\mathcal{C}(y|x,\Theta)}[\log p(x|y, \Phi)] + \mathrm{KL}(q_\mathcal{C}(y|x, \Theta) || p(y)) \quad (1)$$

In **Eq.1**, the first term, i.e., $-\mathbb{E}_{q_\mathcal{C}(y|x,\Theta)}[\log p(x|y, \Phi)]$, represents the mean log-likelihood with respect to the decoder $p(x|y, \Phi)$, while the second term characterized by so-called KL-divergence acts as a regularizer and encourages the decoder $p(x|y, \Phi)$ to be close to a chosen prior distribution $p(y)$. Both $p(x|y, \Phi)$ and $q_\mathcal{C}(y|x, \Theta)$ are systematically treated with deep ANNs in our implementation, as detailed in **Methods** and **Supplementary Note 1 and 4**. To find the optimal settings of $\mathcal{C}$ and $\Phi$, so-called alternatively iterative approach is applied to determine the optimal settings of $\mathcal{C}$ and $\Phi$. In particular, starting with some initializations of $\mathcal{C}$ and $\Phi$, we can calculate $\mathcal{C}$ (resp. $\Phi$,) for $\Phi$ (resp. $\mathcal{C}$) updated in the last iteration step, followed by calculating $\Phi$ (resp. $\mathcal{C}$) based on obtained $\mathcal{C}$ (resp. $\Phi$), repeat this procedure till arriving at some stopping criterion. Note that the optimization with respect to continuous-value variables $\Phi$ and $\Theta$ can be well resolved by calling for back-propagation (BP) optimizers developed in TensorFlow. However, since the element of $c$ has binary value, i.e., 0 or 1, the optimization with respect to them is known to a NP-hard combination optimization problem. In this work, it is heuristically resolved by exploring a randomized simultaneous perturbation stochastic approximation (r-SPSA) algorithm, which is originally developed for treating the problem of optimal well placement and control in the area of petroleum engineering[22]. More details about Eq. 1 and its optimization algorithm implementation can be found in **Supplementary Note 2**.

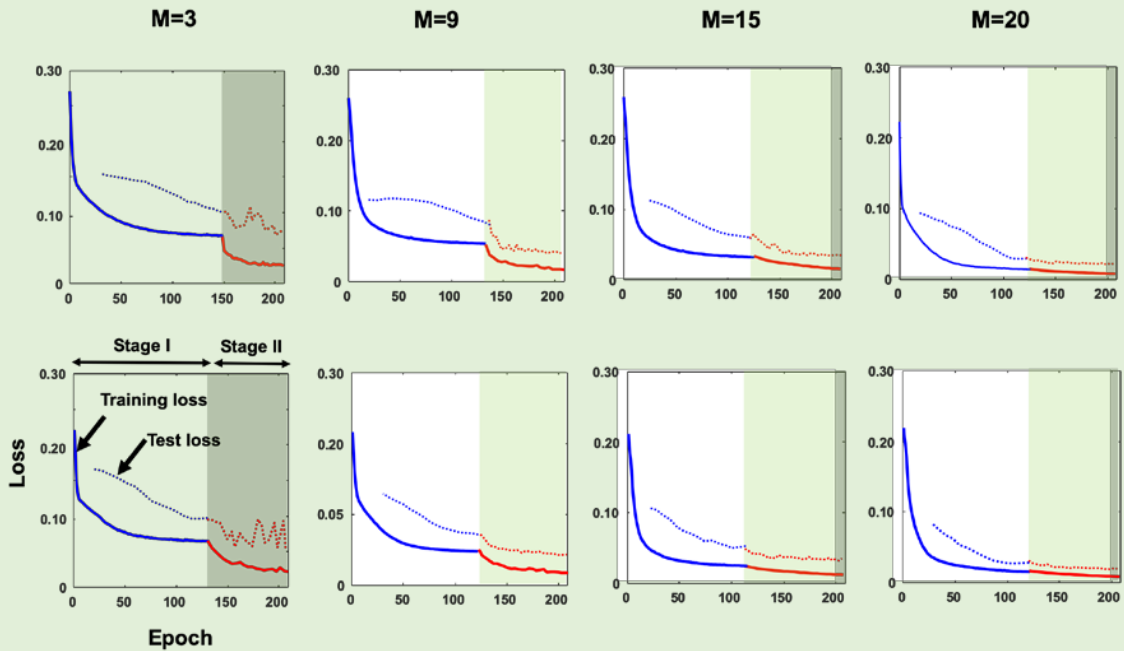

**Figure 2. Working principle of the intelligent sensing system.** (**Top**) The dependence of the training and test loss functions on the progress of iterative epochs for the different number of coding patterns M, i.e., M=3, 9, 15 and 20, where the dashed lines are for the training loss, and the dashed lines are for the test loss. In these figures, the control coding patterns of the metasurface are randomly initialized. Moreover, *stage I* represents the optimization of the r-ANN alone, and *stage II* means the joint optimization of the r-ANN and m-ANN. (**Bottom**) corresponding results with the PCA-based initial coding patterns of metasurface. These results clearly show that the remarkable improvement on the image quality can be achieved by using the joint optimization of the r-ANN and m-ANN, compared to that with the optimization of r-ANN alone, especially when the measurements are highly limited.

**In-situ microwave imaging of human body from reduced measurements.** We here demonstrate the performance of our intelligent sensing system with the in-situ high-resolution imaging of human body in our lab environment. As emphasized previously, we integrate m-ANN for data acquisition and r-ANN for smart data processing as a whole data-driven learnable sensing chain. To that end, we need to jointly optimize the coding patterns of m-ANN together with the weights of r-ANN for the task of human body imaging from microwave raw data. Apparently, the integrated ANN works in a nonlinear ML manner, which can be learned with a standard supervised learning procedure in TensorFlow. In order to demonstrate the significance of the proposed intelligent sensing strategy on the image quality over the conventional learning-based sensing methods, we consider a two-stage training procedure. At the first stage, the coding patterns of m-ANN or the weights of r-ANN are optimized alone. However, at the second stage, m-ANN and r-ANN are *jointly* trained to achieve the overall optimal sensing performance. Additionally, when r-ANN is trained alone at the first stage, the coding patterns of m-ANN are assigned with those generated by two representative linear ML techniques, i.e., random projection and principle component analysis (PCA). In this way, the benefit reaped by the proposed sensing strategy over the conventional methods can be clearly demonstrated. In our study, we use several people, called as training person in short, to train our intelligent microwave sensing system, and use a different person, called as test person, to test it. The details of the training people are provided in Ref. 39.

**Figure 2** displays the cross-validation errors with the progress of iterations for different numbers M of coding patterns on metasurface, i.e., 3, 9, 15, and 20, where the aforementioned two training stages have been distinguished clearly. In order to assess the image quality quantitatively, the so-called structure similarity index metric (SSIM in short) is considered. **Supplementary Figure 4** plots the statistical SSIMs as a function of M for different sensing methods. Since the trainable parameters of m-ANN and r-ANN, except PCA-based m-ANN, are initialized randomly before training, we have conducted 500 realizations and correspondingly made the average over them in order to factor out the effect on the image quality from the choices of random initializations of m-ANN and r-

ANN. The corresponding optimized coding patterns of the metasurface have been reported in **Supplementary Figure 5**.

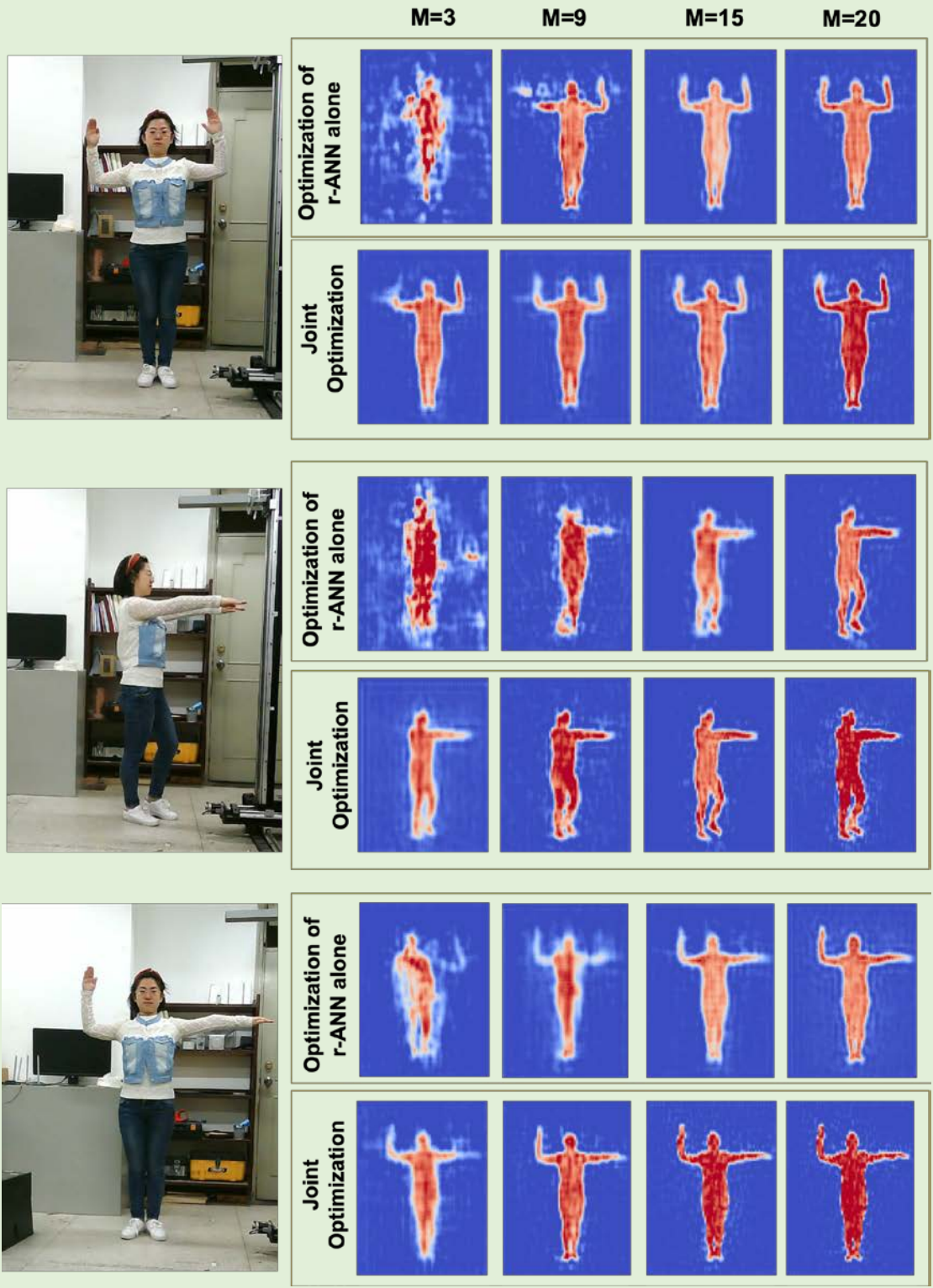

**Figure 3.** **Three sets of experimental results of in-situ microwave imaging from a reduced number of measurements.** For each case, the results in the first row corresponds to the case with the optimization of the r-ANN alone, while the results in the second row for the joint optimization of the r-ANN and m-ANN. Moreover, M denotes the number of coding patterns of metasurface. As predicted above, we can see the remarkable improvement on the image quality by using the joint optimization of the r-ANN and m-ANN, compared to that with the optimization of r-ANN alone, when the number of coding patterns of metasurface is less than 9.

**Figure 3** reports several selected reconstruction results of the test person with different body gestures using the aforementioned sensing methods, where the corresponding coding patterns of the metasurface are displayed in **Supplementary Figure 4**. From these figures, we interestingly find that the image qualities by the joint optimization of r-ANN and m-ANN are much better than those by the optimization of r-ANN alone, regardless of the initial solutions, either PCA or random. This does make sense since more trainable degrees of freedom are involved in the sensing scheme of the joint optimization.

From above results, several important conclusions can be achieved. Overall, the cross-validation error gap between the learned r-ANN (or the learned m-ANN) alone, regardless of the choice of m-ANN (or the learned r-ANN), and the jointly learned setting of r-ANN and m-ANN supports our claim that the proposed intelligent sensing strategy with the simultaneous learning of measurement and reconstruction is remarkably superior to the conventional sensing strategies with optimization of reconstruction or measurement alone, especially when the number of measurements is highly limited. Sharply different from PCA or random based sensing strategies, the presented sensing method can achieve significant improvement on the image quality with remarkable dimensionality reduction by integrating almost all available prior knowledge about the probed scene and the real constraints on measurement setting and processing pipeline into the entire sensing chain.

**In-situ recognition of body gestures from compressed measurements.** Here, we demonstrate the performance of our intelligent sensing system in directly recognizing the body gestures from highly reduced measurements. Similar to above, the m-ANN and r-ANN classifier are merged into as a whole integrated deep ANN, which is simultaneously trained by using a standard supervised learning technique to maximize the whole system's classification accuracy. Similarly, we still use persons with different gestures to train our intelligent sensor, and use a test people to test it. We still consider two-stage training procedure outlined before, where the classical ML technique of random projection is considered for initializing the coding patterns of m-ANN at the first stage.

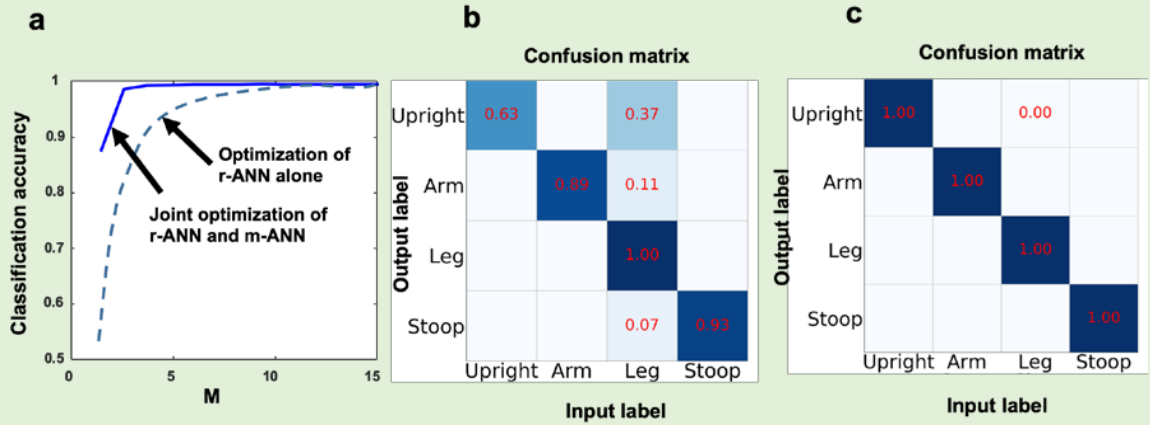

**Figure 4.** Selected experimental results of in-situ gesture recognition directly from a reduced number of measurements. (a) the dependence of the recognition accuracy on the number of coding patterns of the programmable metasurface. (b) The classification confusion matrix with the optimization of the r-ANN alone, where M=3. (c) The classification confusion matrix with the joint optimization of the r-ANN and m-ANN, where M=3.

**Figure 4a** reports the statistical classification accuracy, along with classification error bars, as a function of the number of measurements $M$, where the classification accuracies have been obtained over three different body gestures, and each gesture with 100 samples. Additionally, the corresponding classification matrix with different sensing methods have been reported in **Figure 4b** and **c**. The classification matrix specifically show how often a given body gesture is mistakenly classified by the ANN as one of other gesture options. Note that the strong diagonal elements (corresponding to correctly identified body gestures) reflect the achieve recognition accuracy of almost 100% on average by using the intelligent sensing method. The performance using our intelligent sensing settings saturates around M=5 at 98%, meaning that we can manage to extract task-relevant information with only 3 coding patterns of metasurface. For M≤5, our scheme yields the gain in accuracy of the order of 5~35% which is a substantial improvement in the context of classification tasks. Above results verify our claim that, for the task of recognizing human-body gestures from highly reduced measurements, the proposed method with the simultaneous optimization of measurement and reconstruction procedures will remarkably outperform conventional strategies with optimization of reconstruction or measurement alone. Such sensing paradigm is apparently helpful since taking fewer measurements takes less time and less energy consumption.

**Discussions and Conclusions**

We proposed a data-driven learnable EM sensing paradigm by integrating a programmable metasurface into the data processing pipeline as a whole. The integrated sensing chain is composed of two dedicated deep ANNs, i.e., m-ANN for smart data acquisition and r-ANN for instant data processing. Then, the sensing strategy with optimal sensing performance can be achieved by jointly training m-ANN along with matching r-ANN in sense of VAE. As such, the presented sensing methodology can account for almost all available prior knowledge about the probed scene and real constraints on the measurement setting and processing pipeline in the entire sensing chain. As a result, the proposed sensing strategy with simultaneous learning of measurement and reconstruction procedures is superior to the conventional sensing strategies with the optimization of reconstruction or measurement setting alone, especially when the number of measurements is highly limited, as we have experimentally demonstrated. Moreover, an interesting insight from our results is that the learned measurement setting, i.e., the illuminations generated by the optimized coding patterns of the programmable metasurface, perform better than those with generic random illuminations or PCA-based measurement modes.

To summarize, we have discussed how *artificial material* and *artificial intelligence* are merged in the design of intelligent sensing architecture, and demonstrated how the data-driven learnable intelligent sensing system can be leveraged across the whole sensing chain. The presented methodology could offer a fundamentally new perspective on the design of sensing architectures at various frequencies, and beyond. We faithfully expect that all these results would initiate a new line of research towards building efficient intelligent sensing architectures.

## Methods

**Design of programmable coding metasurface.** The programmable metasurface, as an emerging active member of artificial material family[22-31], is an ultrathin planar array of electronically-controlled meta-atoms[32-44]. Owing to the unique capability in manipulating EM wavefields in a reprogrammable way, it has elicited many exciting physical phenomena and versatile functional devices, such as computational imager[37-40]. programmable holography[35], wireless communication system[41-43], and others[34, 44, 45].

The designed programmable metasurface consists of 32x24 meta-atoms operating at around 2.4 GHz, as shown in **Supplementary Figure 1**, in which the meta-atom with size of 54mm×54mm are detailed. The designed meta-atom is composed of two substrate layers: the top layer is F4B with the relative permittivity of 2.55 and loss tangent of 0.0019; and the bottom is FR4 with size of $0.54 \times 0.54$ mm$^2$. The top square metal patch, with size of $0.37 \times 0.37$ mm$^2$, is integrated with a PIN diode (SMP1345-079LF) to control the EM reflection phase of meta atom. In addition, a Murata LQW04AN10NH00 inductor with inductance of 33nH is used to achieve the good isolation between the RF and DC signals. In our design, the whole programmable metasurface is composed of 3×3 metasurface panels, and each panel is composed of 8×8 electronically-controllable digital meta atoms. Each metasurface panel is equipped with eight 8- bit shift registers (SN74LV595APW), and eight PIN diodes will be sequentially controlled. The CLK rate adopted is 50MHz, and the ideal switching time of PIN diodes in is 10us.

**Model of the m-ANN.** We here elaborate on the details about the measurement ANN (m-ANN) for the data-driven learnable data acquisition based on the designed programmable metasurface. Conventionally, an optimal task-adaptive measurement strategy is hardly to be determined in a tractable way, especially when the hardware settings or/and surrounding environments are so complex that they cannot be well described. With advent of deep learning techniques, this open challenging problem can be efficiently resolved using deep learning techniques.

Unlike conventional end-to-end deep ANNs that have two ports, the m-ANN presented here has three ports: one is for receiving the image or label of probed scene $x$, one is for the coding pattern of the metasurface $\Theta$, and the other is for outputting the measurements (i.e., the microwave raw data in this work) $y$, as illustrated in **Figure 1**. In a deterministic standpoint, we can describe the three-port m-ANN with a pair of coupled nonlinear equations, i.e.,

$$y = f(x; \mathcal{W}_\mathcal{C}) + n \tag{2}$$

and
$$\mathcal{W}_\mathcal{C} = g(\mathcal{C}; \Theta) + N \tag{3}$$

Herein, $n$ and $N$ involved in Eqs. (2)-(3) represent their own modelling errors, whose elements are considered as i.i.d. complex-valued Gaussian random numbers. In **Eq. 2**, $\mathcal{W}_\mathcal{C}$ encapsulates all learnable weights of an end-to-end ANN relating the input $x$ to the desired measurement $y$. $\mathcal{W}_\mathcal{C}$ imposed with the subscript $\mathcal{C}$ highlights a fundamental fact that these trainable network weights depend on the coding patterns of metasurface $\mathcal{C}$ through **Eq. 3**. In our implementation, these two nonlinear equations, i.e., $f$ and $g$, are modeled with deep convolutional neural networks with trainable weights $\mathcal{W}_\mathcal{C}$ and $\Theta$, respectively. Both $\mathcal{W}_\mathcal{C}$ and $\Theta$ can be readily learned with a standard supervised learning procedure from triplet training samples $\{x^{(i)}, \Theta^{(i)}, y^{(i)}, i = 1,2, ...\}$, where the superscript denotes the index of training samples. More details about them can be found in **Supplementary Note 4**.

Note that after the m-ANN is well trained, the coding patterns of metasurface can be readily learned to match with the r-ANN, as described previously. Furthermore, as outlined above, the output from the m-ANN, the measurement vector, is injected into the learnable r-ANN. The final digital layer's output is the image or classification of the scene of interest.

By jointly training the coding patterns of m-ANN and the weights of r-ANN, we identify measurement settings that optimally match the constraints and digital processing layer.

**Training of the m-ANN and r-ANN.** The intelligent sensing system consists of two deep ANNs, i.e., the m-ANN and r-ANN. The m-ANN is responsible for the adaptive data acquisition on the physical level, while the r-ANN is for the instant data processing on the digital pipeline.

As for the training of the complex-valued weights of m-ANN and r-ANN, the training stage is performed by the ADAM optimization method with the mini-batch size of 32 and epoch of 101. The learning rates are set to $10^{-4}$ and $10^{-5}$ for the first two layers and the last layer, and halved once the error plateaus. The complex-valued weights are initialized by random weights with zero-mean Gaussian distribution and standard deviation of $10^{-3}$. The trainings are performed on a workstation with Intel Xeon E5-1620v2 central processing unit, NVIDIA GeForce GTX 1080Ti, and 128GB access memory. The machine learning platform Tensor Flow[46] is used to design and train the networks in the intelligent metasurface system.

**Configuration of proof-of-concept sensing system.** We build a proof-of-principle intelligent system in microwave. The experimental setup consists of a transmitting (Tx) horn antenna, a receiving (Rx) horn antenna, a large-aperture programmable metasurface, and a vector network analyzer (VNA, Agilent E5071C). In our sensing experiments, two horn antennas are connected to two ports of VNA through two 4m-long 50-Ω coaxial cables, and the VNA is used to acquire the response data by measuring transmission coefficients ($S_{21}$). To suppress the measurement noise level, the average number and filtering bandwidth in VNA are set to 10 and 10 kHz, respectively. More details can be found in **Supplementary Note 1**.

## Acknowledgments

This work was supported in part from the National Key Research and Development Program of China (2017YFA0700201, 2017YFA0700202, and 2017YFA0700203), and in part from the National Natural Science Foundation of China (61471006, 61631007 and 61571117).

## Author Contribution

L.L. conceived the idea, conducted theoretical analysis and wrote this paper. H.-Y. L. conducted experiments and data processing. All authors participated in the experiments, data analysis, and read the manuscript.

## Additional Information

**Supplementary Information** accompanies this article.

**Competing Interests**: The authors declare no competing financial interests.